\documentclass[pdflatex]{pasj02} 
\usepackage[switch,mathlines]{lineno} 
\usepackage{fancyvrb}
\usepackage{graphicx}
\usepackage{natbib}

\jyear{2026}
\Received{}
\Accepted{}

\begin{document} 
\VerbatimFootnotes

\title{ 
Jet-torus interaction revealed by sub-parsec SO absorption in NGC 1052
}

\author{
Satoko \textsc{Sawada-Satoh},\altaffilmark{1,2}
\altemailmark\orcid{0000-0001-7719-274X}\email{swdsth@gmail.com} 
 Seiji \textsc{Kameno},\altaffilmark{3,4,5}\orcid{0000-0002-5158-0063}
 Nozomu \textsc{Kawakatu},\altaffilmark{6}\orcid{0000-0003-2535-5513}
 Do-Young \textsc{Byun},\altaffilmark{7,8}
Se-Jin \textsc{Oh},\altaffilmark{7}
Sang-Sung \textsc{Lee},\altaffilmark{7,8}
Duk-Gyoo \textsc{Roh},\altaffilmark{7}
Chungsik \textsc{Oh},\altaffilmark{7}
Jae-Hwan \textsc{Yeom},\altaffilmark{7}
Dong-Kyu \textsc{Jung},\altaffilmark{7}
Hyo-Ryoung \textsc{Kim},\altaffilmark{7}
Young-Sik \textsc{Kim},\altaffilmark{7} 
and 
Sanghyun \textsc{Kim}\altaffilmark{7}
}

\altaffiltext{1}{Graduate School of Science, Osaka Metropolitan University, 3-3-138 Sugimoto Sumiyoshi-ku, Osaka 558-8585, Japan
}
\altaffiltext{2}{Faculty of Engineering, Fukui University of Technology, 
3-6-1 Gakuen, Fukui City, Fukui 
910-8505, Japan
}
\altaffiltext{3}{Joint ALMA Observatory, Alonso de \'{o}rdova 3107 Vitacura, Santiago 763-0355, Chile}
\altaffiltext{4}{National Astronomical Observatory of Japan, 2-21-1 Osawa, Mitaka, Tokyo 181-8588, Japan}
\altaffiltext{5}{Department of Astronomy, School of Science, Graduate University for Advanced Studies (SOKENDAI), Tokyo 181-8588, Japan}
\altaffiltext{6}{National Institute of Technology, Kure College, 2-2-11 Agaminami, Kure, Hiroshima 737-8506, Japan}
\altaffiltext{7}{Korea Astronomy and Space Science Institute, 776 Daedeok-daero, Yuseong, Daejeon 34055, Republic of Korea}
\altaffiltext{8}{University of Science and Technology, 217 Gajeong-ro, Yuseong-gu, Daejeon 34113, Republic of Korea}



\maketitle

\begin{abstract}
We report the first $\lambda$2-mm very long baseline interferometry (VLBI) observations of the radio galaxy NGC~1052, conducted with the Korean VLBI Network (KVN) using a wide-band recording mode. 
Leveraging the wide bandwidth covering a velocity range at 2300 km~s$^{-1}$,  
we successfully detect broad ($> 700$ km~s$^{-1}$) multi-component SO $J_N = 3_3-2_2$ absorption  
against the sub-parsec-scale continuum structure. 
The absorption profile consists of 
both redshifted and blueshifted components, 
including a newly identified blueshifted feature 
at $-412$ km~s$^{-1}$ relative to the systemic velocity. 
Significant SO absorption is confined to the central components, 
with no substantial detection toward the outer jet components. 
This constrains the location of SO gas to a compact region smaller than 0.45 pc in the sub-parsec vicinity 
of the supermassive black hole (SMBH). 
Our results 
support the scenario in which
SO molecules are evaporated through shock heating caused by jet-torus interaction. 
The SO gas clumps are likely driven outward 
by the jet, with some returning toward the SMBH as inflowing material.
Comparison with 321 GHz H$_2$O masers reveals partial similarities 
in spatial distribution and radial velocity, 
suggesting that 
the jet-torus interaction may also trigger the excitation of H$_2$O masers. 
\end{abstract}


\section{Introduction}

The properties of circumnuclear media at the centers of galaxies 
play a crucial role in various phenomena associated with active galactic nuclei (AGNs). 
It is widely accepted that AGNs are powered 
by the accretion of circumnuclear gas
\citep{lynden69}, 
which serves as the reservoir fueling the central supermassive black hole (SMBH). 
Moreover, the circumnuclear medium is a key component of the AGN unification model. 
A toroidal distribution of gas surrounding the SMBH, 
commonly referred to as the ``torus'',
is responsible for
obscuration of the broad-line region, 
thereby naturally explaining the observed dichotomy between type 1 and type 2 Seyfert galaxies \citep{antonucci93, urry95}. 
In addition, 
recent observations of nearby AGN jets  
using  Very Long Baseline Interferometry (VLBI) 
have revealed that 
the relativistic jets are highly 
collimated within the nuclear region 
\citep[e.g.,][]{asada12,hada13}. 
These findings suggest that the circumnuclear medium may play a significant role in collimating jets by exerting pressure. 
Therefore, 
detailed investigations into the morphology and kinematics of
 the circumnuclear gas surrounding the SMBH are essential for advancing our understanding of AGN fueling mechanisms and jet collimation processes.


NGC 1052 is a nearby elliptical galaxy 
with a systemic velocity (V$_\textrm{sys}$) of $1492 \pm 2.6$ km~s$^{-1}$ \citep{kameno20} 
with respect to the local standard of rest (LSR). 
Its central region hosts a low-luminosity AGN 
characterized by a low-ionization nuclear emission-line region (LINER) spectrum  \citep[e.g.,][]{gabel00}, 
as well as a hidden broad line region (BLR) detected in polarized light 
\citep{barth99}. 
The nucleus is surrounded by a central condensation of ambient gas in multiple phases. 
Past high-resolution observations have found 
atomic and molecular absorption lines 
including 
H~\textsc{i} \citep{vermeulen03}, 
OH \citep{omar02, impellizzeri08}, 
CO, HCN, HCO$^{+}$, SO, SO$_2$, CS, CN, 
and H$_2$O 
toward the center of NGC~1052
 \citep{liszt04, kameno20}.  
H$_2$O megamaser emission has also been mapped 
in the parsec (pc) scale region 
using the Very Long Baseline Array (VLBA) at 22 GHz \citep{claussen98, sss08}
and 
 the Atacama Large Millimeter/submillimeter Array 
(ALMA) at 321 GHz 
\citep{kameno23b,kameno24}. 
Multi-frequency VLBI images have also shown 
a central condensation of ionized gas 
surrounding the SMBH
\citep{kameno01,kameno03}. 
This multiphase gas condensation has been 
interpreted as 
a circumnuclear torus 
comprising several layers with temperatures ranging from $10^2$ to $10^4$ K 
\citep{kameno05,sss08}. 
Hot ($10^4$ K) ionized gas resides
on the inner surface layer and obscures the base of the nuclear jet, producing the emission gap between the eastern approaching  and western receding jets at frequencies of $\le$ 22 GHz 
due to free--free absorption (FFA) by ionized gas 
\citep{kameno01,kameno03,vermeulen03,kadler04,sss08}. 
A nuclear component is obscured within this  emission gap below 43 GHz, 
but becomes visible at 43 GHz and higher frequency bands 
\citep{kadler04,sss08,baczko16,sss16,sss19}. 
Adjacent to the ionized layer,
a warm ($\sim$ 400 K) molecular layer  
contains excited H$_2$O molecules 
with an inverted population distribution,  
that produces the 22 GHz maser emission.  
Cooler molecular gas ($< 400$ K) 
lies at outer radii of the torus,  
and molecular absorption lines can be detected 
on the background continuum component. 
The gas in the molecular layers exhibits redshifted spectral lines \citep{sss08,kameno23a} indicating infall motion toward the SMBH.
ALMA observations of 
 molecular absorption lines by \citet{kameno20} 
proposed a thick torus geometry 
with a radius of 2.4$\pm$1.3 pc 
and a height-to-radius ratio of 0.7$\pm$0.3. 
Modeling studies indicate that 
the torus models with a steep density gradient 
along the jet axis can 
simultaneously reproduce 
 the sub-parsec (sub-pc) scale distribution of FFA opacity and the broadband X-ray spectra 
\citep{balokovic21}. 

NGC 1052 is also known to host a prominent double-sided radio jet ranging from sub-pc to kilo-pc scales along the east-west direction \citep[e.g.,][]{jones84, wrobel84, kellermann98}. 
The jet maintains a well-collimated structure 
within a distance of $10^4$ Schwarzschild radii 
from the SMBH \citep{nakahara20, baczko22}, 
whose mass is estimated to be $10^{8.19}$ $M_{\odot}$ 
\citep{woo02}
. 
It has been suggested that the nuclear jet interacts with dense gas clumps embedded in a geometrically thick torus, 
and 
such jet-torus interaction 
can lead to the jet collimation on sub-pc scale 
 \citep[e.g.,][]{fromm19}. 
ALMA observations of the 321 GHz H$_2$O maser emission 
in NGC 1052 
further indicate that 
such jet-torus interactions can also affect 
the excitation of the H$_2$O masers  
\citep{kameno23b,kameno24}. 
Observations of sulfur-bearing species 
such as SO, SO$_2$ and CS 
are particularly valuable for probing the jet–torus interaction,  
as these species are considered to be reliable shock tracers 
 \citep[e.g.,][]{pineau93}.

ALMA studies of sulfur-bearing molecular absorption 
have revealed a warm ($344\pm43$ K) environment  
in the torus, 
derived from submillimeter SO absorption lines 
at rest frequencies ($\nu_{\rm rest}$) of $\ge$ 240 GHz 
\citep{kameno20, kameno23a}. 
This temperature is consistent with 
the presence of 22 GHz H$_2$O maser emission, and 
the detection of 
vibrationally excited HCN and HCO$^{+}$ absorption lines,  
both of which are associated with the torus.
\citet{kameno23a} suggested that 
the jet–torus interaction is responsible for shock heating of gas and dust 
in the inner regions of the torus, 
where SO molecules are likely to desorb from the icy mantles of dust grains. 
Moreover, 
the observations of millimeter-wave SO absorption lines at rest frequencies 
 $\nu_{\rm rest}$ of $\le$ 129 GHz 
reveal complex and asymmetric line profiles, 
including a redshifted peak 
around $V_{\rm sys} + 160$ km~s$^{-1}$, 
a sharp redward edge
around $V_{\rm sys} + 200$ km~s$^{-1}$, 
and a shallow blueward slope down to 
$V_{\rm sys} - 200$ km~s$^{-1}$.  
These features point to a dynamic and inhomogeneous medium comprising multiple kinematic components and varying line-of-sight geometries relative to the background continuum.
To gain deeper insight into the jet-torus interaction, 
it is crucial to spatially resolve the millimeter SO absorption and to precisely determine both 
the spatial distribution and kinematic properties of the SO gas through high-resolution observations.

Following the ALMA studies that detected jet–torus interaction by \citet{kameno23a}, 
we carried out VLBI observations of the 
 SO $J_N = 3_3-2_2$ absorption line  
($\nu_{\rm rest} =$  129.13892 GHz) in the sub-pc 
region of NGC 1052 
using the Korean VLBI Network (KVN) 
in wideband mode. 
The SO $J_N = 3_3-2_2$ absorption line was previously detected with a peak optical depth of 0.0214 
using  ALMA \citep{kameno23a}. 
We adopt a redshift of $z =$ 0.005 for NGC~1052, 
such that 
one milliarcsecond (mas) corresponds to 0.095 pc 
in the galaxy.
The radial velocities are hereafter expressed relative to the LSR.

\section{Observations and data reduction}

KVN Observations of NGC 1052 were conducted 
on 2020 January 4 and 5, 
yielding a total on-source time of 14.5 hr. 
In addition to NGC 1052, the bright continuum sources 
3C~84, NRAO~150 and J0423$-$0120 were observed 
for 6 minutes every 1 hr 
for fringe finding and calibration. 
To improve the sensitivity for the $\lambda$2 mm 
SO $J_N = 3_3-2_2$ absorption line observations, 
simultaneous dual-frequency data at 21.5 GHz ($\lambda$1.3 cm)
and 129 GHz ($\lambda$2 mm) were recorded 
using the KVN multi-frequency receiving system 
\citep{han08,han13} 
and the new high-speed sampler OCTAD 
\citep{oyama16, oyama24}. 
The data were acquired at a rate of 16 Gbps 
with four spectral windows (SWs), 
each covering 1024 MHz 
with dual circular polarization.

Two of four SWs were assigned to 
left- and right-hand circular polarization (LHCP and RHCP) 
at 129 GHz 
for the target frequency band. 
The remaining two SWs were set to 
LHCP and RHCP at 21.5 GHz  
for the phase referencing. 
The velocity coverage of one SW 
was 2300 km~s$^{-1}$ at 129 GHz. 
Correlation was carried out 
with the DiFX software correlator \citep{deller07} 
at the Korea-Japan Correlation Center 
(\cite{lee15}; 
see also KVN Status Report\footnote{\verb|https://radio.kasi.re.kr/status_report/files/KVN_status_report_2025.pdf|}), 
outputting two parallel-hand polarization (LL and RR) 
visibility products. 
The parallel-hand polarization data were later 
averaged into the total intensity data 
during post-correlation processing.

Post-correlation processing, 
including calibration, data editing 
and imaging, was performed  
using the NRAO Astronomical Image Processing System  \citep[AIPS;][]{greisen03}. 
Since the KVN antennas have altitude-azimuth mounts, parallactic angle correction was applied at the outset. 
A priori amplitude calibration was conducted using elevation-dependent gain curves and system temperature measurements recorded at each station during the observations. 
Corrections for Doppler-shifted velocities in the SO absorption line, caused by Earth’s motion, were also applied. 
Complex bandpass calibration for each circular polarization hand was performed 
using 3C~84, NRAO~150 and J0423$-$0120. 
The amplitude and phase components of the bandpass solutions were derived from auto-correlation and cross-correlation data, respectively 
\citep[e.g.,][]{kemball95}. 
Prior to applying the bandpass correction, residual delays and rates on the bandpass calibrators were removed to ensure coherent visibility phases across frequency and time. 
We then derived the complex bandpass solutions by coherently averaging the data over time. 
The resulting bandpass solutions were applied to the data, yielding residual amplitude ripples corresponding to a fractional amplitude error of less than 2~\%. 
To mitigate rapid phase fluctuations at 129 GHz caused by atmospheric turbulence, we employed the frequency phase transfer method, transferring phase solutions from the lower-frequency data (21.5 GHz) to the higher-frequency data (129 GHz) by scaling according to their frequency ratio \citep{middelberg05, rioja11}. 

We generated continuum data by averaging all channels,
and a spectral cube with a frequency interval 
of 16 MHz. 
Fringe-fitting and self-calibration solutions derived from the continuum data were applied to the cube.
The visibility data were imaged without $uv$-taper 
using natural weighting. 
The resulting synthesized beam size 
is 0.88 $\times$ 0.73 mas, 
corresponding to 0.084 $\times$ 0.069 pc in NGC 1052. 
Continuum emission was identified and subtracted from the spectral cube by fitting a polynomial to line-free channels.

Finally, 
optical depth images of SO absorption were derived by combining the continuum map and the line channel maps. 
We clipped out image pixels with 
intensities below 7.1 mJy beam$^{-1}$ 
($<$ 5 $\sigma$) 
in the continuum map 
due to poor signal-to-noise ratio in the optical depth 
at positions with weak continuum emission.
We note that 
the SO $J_N = 3_3-2_2$ line 
($\nu_{\rm rest} =$  129.13892 GHz)
is close in frequency 
to SO$_2$ $J_{Ka,Kc} = 12_{1,11}-11_{2,10}$ line 
($\nu_{\rm rest} =$  129.10583 GHz)  
and that contamination from 
the SO$_2$ $J_{Ka,Kc} = 12_{1,11}-11_{2,10}$ line is therefore  possible. 
The SO$_2$ $J_{Ka,Kc} = 10_{2,8}-10_{1,9}$ 
($\nu_{\rm rest} =$ 129.51481 GHz)
and $J_{Ka,Kc} = 12_{2,10}-12_{1,1}$ 
($\nu_{\rm rest} =$ 128.60513 GHz) transitions  
at adjacent frequencies exhibit $\tau_{\rm max} = 0.0060 \pm 0.0008$ and $0.0059 \pm 0.0008$, respectively, while $\tau_{\rm max} $ of SO $J_N = 3_3 - 2_2$ is $0.0214 \pm 0.0012$ \citep{kameno20}. Thus, the contamination of the SO$_2$ $J_{Ka,Kc} = 12_{1,11}-11_{2,10}$ absorption would account for approximately $28$\%.

\section{Results}

\subsection{Continuum nuclear emission}

\begin{figure}[tb]
 \begin{center}
  \includegraphics[width=8cm]{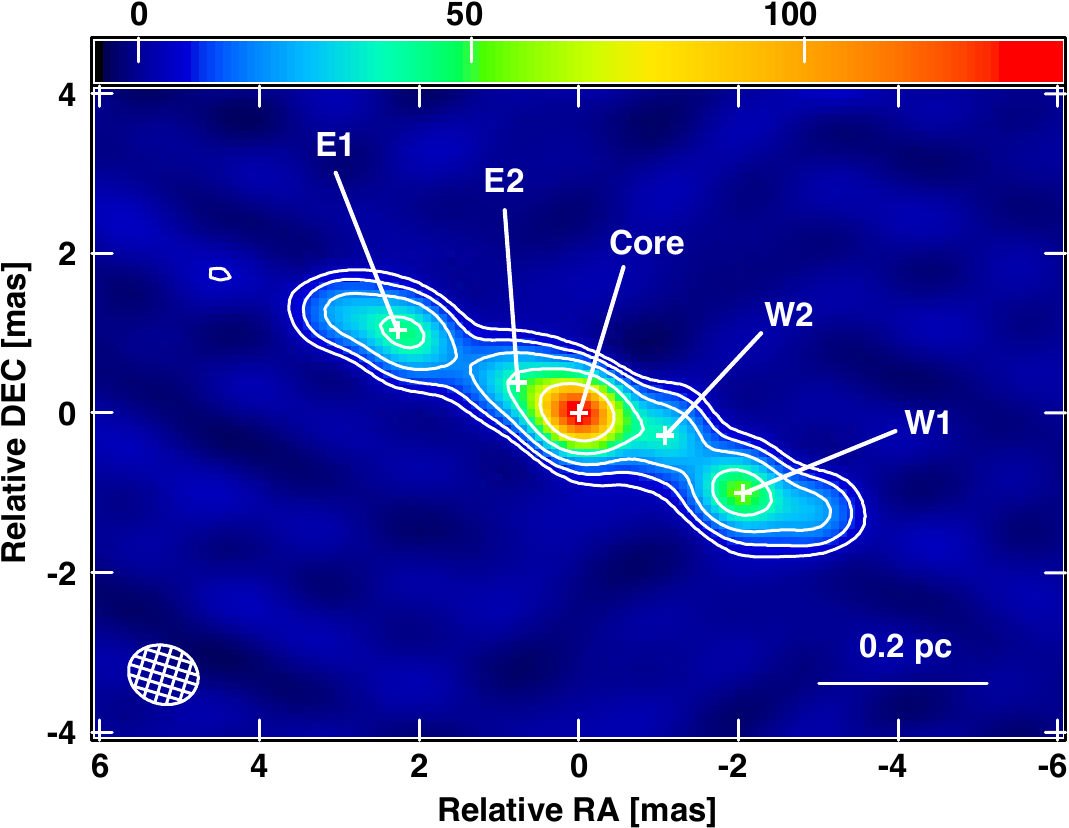} 
 \end{center}
\caption{
Continuum map of the nuclear region 
in NGC 1052 at 129 GHz. 
Contours begin at 3 times $I_{\rm rms}$
and increase by a factor of 2, 
where $I_{\rm rms}$ = 1.42 mJy~beam$^{-1}$. 
The peak intensity is 125 mJy~beam$^{-1}$. 
The synthesized beam is 0.88 $\times$ 0.73 mas  
at a PA of 73$^{\circ}$ 
as represented by a cross-hatched ellipse 
in the bottom-left corner. 
The plus symbols mark the location of 
fitted Gaussian components E1, E2, Core, W2 and W1. 
The parameters obtained from the Gaussian components are listed 
in table~\ref{tab:cntgauss}.
 {Alt text: 
 Contour map of the sub-pc region of NGC 1052 at 129 GHz, showing five labeled Gaussian components (E1, E2, Core, W2, and W1)
} 
}
\label{fig:dcntmap}
\end{figure}

The 129 GHz continuum image of NGC 1052 
is shown in figure~\ref{fig:dcntmap}. 
This is the first VLBI image at 129 GHz of  
the sub-pc scale continuum jet structure in this galaxy. 
A bright central source and 
a two-sided jet structure, 
with the eastern side approaching and the western side receding, 
extending over 8 mas (0.75 pc), 
are spatially resolved with the KVN synthesized beam.  
The jet appears to be well collimated. 
The bright central source is resolved into 
a bright component and 
 the innermost jet components on both sides.

We identify five continuum components 
by performing two-dimensional Gaussian model fitting 
on the brightness distribution using the AIPS task JMFIT. 
These components are labeled as the eastern outer jet component (E1), the eastern innermost jet component (E2), the brightest component (Core), the western innermost jet component (W2), and the western outer jet component (W1), as shown in figure~\ref{fig:dcntmap}.
The derived Gaussian fit parameters are summarized in table~\ref{tab:cntgauss}.
The continuum components are aligned 
along a position angle of $64^{\circ}\pm2^{\circ}$, 
which is in agreement with past VLBI 
observations at lower frequency bands  
\citep{jones84, kameno01, kadler04}.


\begin{table*}[ht]
  \tbl{Parameters of continuum components by two dimension Gaussian fits}{%
  \begin{tabular}{lccccc}
    \hline \hline
    Component
    & $S$ 
    & $\Delta$R.A. 
    & $\Delta$Decl.
    & Major Axis
    & Minor Axis \\
    & (mJy) & (mas) & (mas) & (mas) & (mas)  \\
    (1) & (2) & (3) & (4) & (5) & (6) \\
      \hline
    E1 & $65\pm3$ & $+2.26\pm0.02$  & $+1.03\pm0.01$ & $1.52\pm0.05$ & $0.72\pm0.02$  \\
    E2 & $45\pm3$ & $+0.76\pm0.02$  & $+0.38\pm0.01$ & $1.19\pm0.05$ & $0.75\pm0.03$  \\
    Core & $110\pm2$ & $0.00\pm0.01$ & $0.00\pm0.01$ & $0.87\pm0.01$  & $0.71\pm0.01$  \\
    W2 & $19\pm2$ & $-1.09\pm0.02$  & $-0.29\pm0.02$ & $0.94\pm0.06$ & $0.59\pm0.04$  \\ 
    W1 & $72\pm3$ & $-2.07\pm0.02$  & $-1.01\pm0.01$ & $1.37\pm0.04$ & $0.74\pm0.02$  \\
    \hline
  \end{tabular}}
\label{tab:cntgauss}
\begin{tabnote} 
Col.(1) Component ID. 
Col.(2) Flux density. 
Cols.(3)--(4) Relative position with respect to the phase center.
Cols.(5)--(6) Full width at half maximum (FWHM) of the Gaussian component. 
Deconvolved major and minor axis sizes. 
\end{tabnote}
\end{table*}

\subsection{SO 
 $J_N = 3_3-2_2$ 
absorption}\label{sec:so}

\begin{figure*}[tb]
 \begin{center}
  \includegraphics[width=16cm]{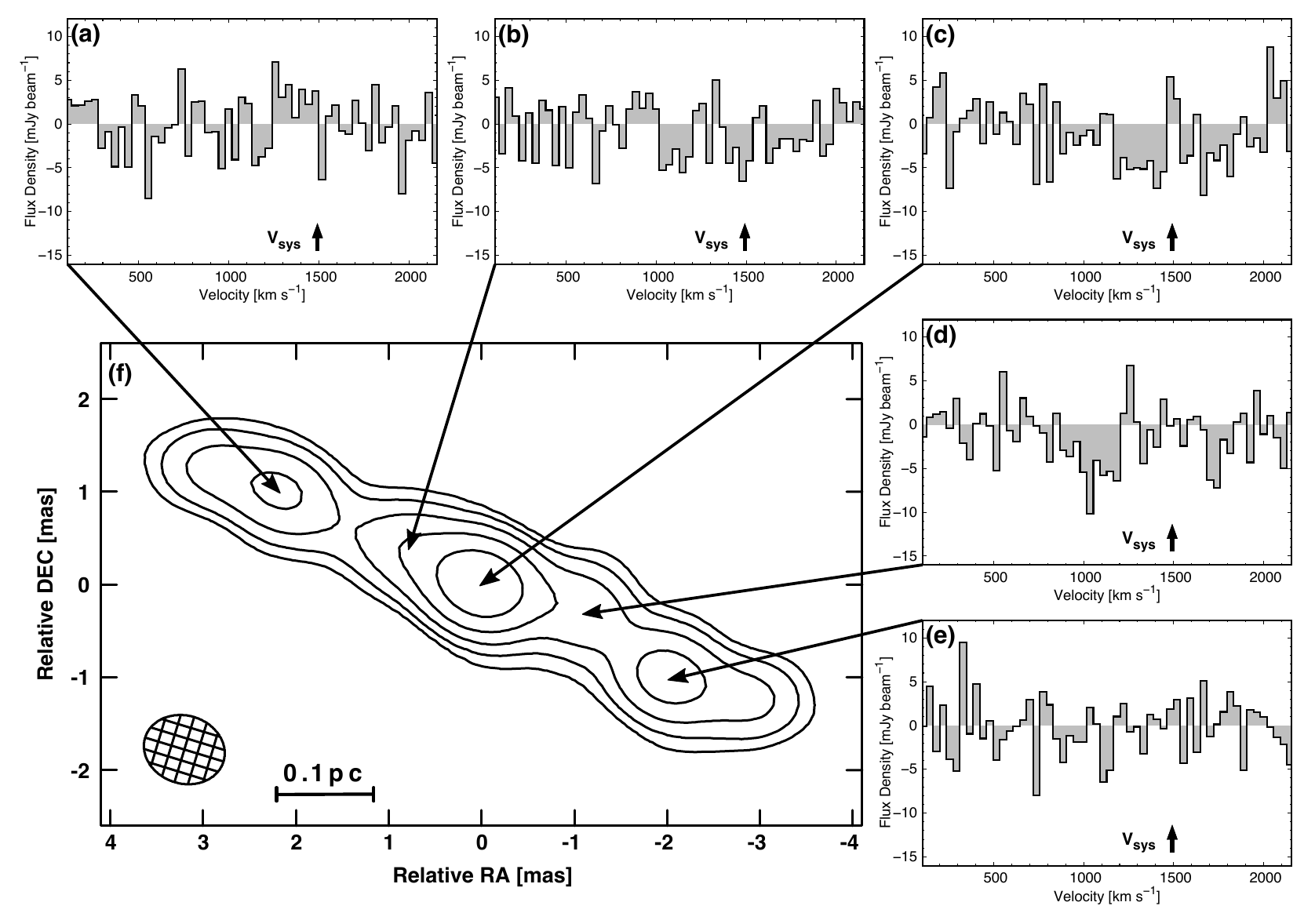} 
 \end{center}
\caption{
(a)--(e) 
SO $J_N = 3_3-2_2$ absorption spectra 
integrated over a  0.3 $\times$ 0.3 mas$^2$ region 
at different locations of E1, E2, Core, W2 and W1, 
as marked by arrows. 
The zero flux level corresponds to the continuum level at each location. 
The velocity resolution is 37.1 km~s$^{-1}$. 
(f) Continuum contour map of NGC 1052 at $\lambda$2 mm also 
shown in figure~\ref{fig:dcntmap}. 
{Alt text:
Five panels showing SO line spectra 
at different locations (E1, E2, Core, W2, and W1),  
along with a continuum map of NGC 1052 
indicating their spatial positions. 
}
}
\label{fig:soispec}
\end{figure*}

Spectral profiles of 
SO $J_N = 3_3-2_2$ absorption 
toward the continuum components (E1, E2, Core, W2, and W1) are presented in figure~\ref{fig:soispec}a--e. 
These profiles were 
extracted from the spectral cube by 
integrating over $0.3\times0.3$ mas$^2$ regions  
at each continuum component. 
The frequency resolution is 16 MHz, 
corresponding to a velocity resolution of 37.1 km~s$^{-1}$.  
The typical rms noise per 16 MHz channel
is 3 mJy~beam$^{-1}$ 
in the cube. 
SO absorption features are clearly detected
on the central components Core and W2, 
and marginally on the component E2  
(figure~\ref{fig:soispec}b--\ref{fig:soispec}d). 
No significant absorption is found 
toward the jet components E1 and W1 
(figure~\ref{fig:soispec}a and \ref{fig:soispec}e).

\begin{figure}
 \begin{center}
  \includegraphics[width=8cm]{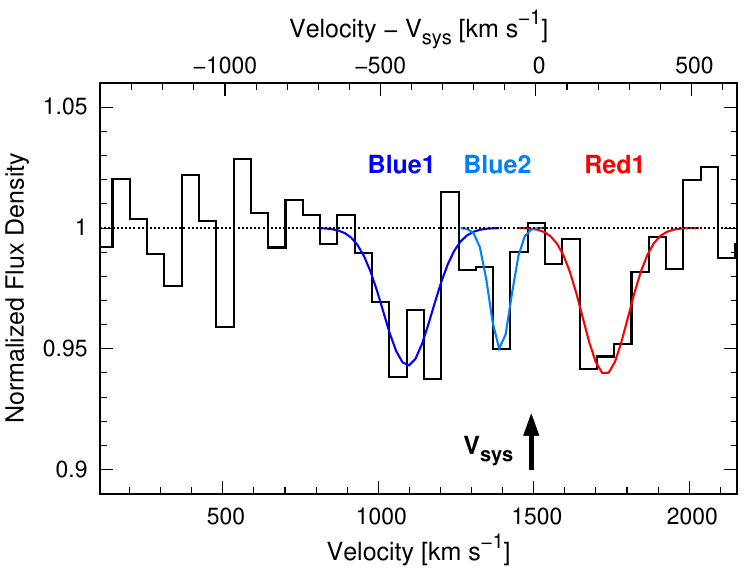} 
 \end{center}
\caption{
Spectral profile of SO absorption 
integrated over 
three 0.3 $\times$ 0.3 mas$^2$ regions centered 
on the continuum components E2, Core, and W2. 
The spectrum is normalized by the combined continuum flux densities 
of E2, Core and W2. 
The velocity resolution is 55.6 km~s$^{-1}$, and 
the rms noise level is 0.017 in normalized flux density. 
The three colored curves indicate 
a triple-Gaussian fit to the observed profile. 
The fitting results are listed in table~\ref{tab:sogauss}. 
{Alt text: 
A line graph displaying the SO absorption spectrum 
with three fitted components 
labeled Blue1, Blue2, and Red1. 
}
}
\label{fig:socenter}
\end{figure}

Figure~\ref{fig:socenter} displays 
the spectral profile of the SO absorption line
integrated over three $0.3\times0.3$ mas$^2$ regions
centered on the components E2, Core, and W2. 
The spectrum is binned to a
velocity resolution of 55.6 km~s$^{-1}$ per channel, and 
all channels are normalized to the continuum level. 
Absorption features exceeding three times 
the rms noise (0.017 in normalized flux) are detected 
at the velocity channels of 1015, 1175, 1397, 1675 and 1731 km~s$^{-1}$. 
The latter two redshifted channels (1675 and 1731 km~s$^{-1}$)  
have velocities close to the HCN absorption peaks 
dominated 
by two redshifted components at 1656 and 1719 km~s$^{-1}$ 
reported by \citet{sss16}. 
The blueshifted feature 
at 1397 km~s$^{-1}$ appears to correspond to the blueshifted component B1 identified in the 321 GHz H$_2$O maser emission by \citet{kameno24}. 
The spectral profile reveals multiple velocity components, confirming both redshifted and blueshifted peaks 
around $V_{\rm sys}+200$ km~s$^{-1}$ 
and $V_{\rm sys}-120$ km~s$^{-1}$, 
as previously identified in the asymmetric profile with ALMA
\Citep[$V_{\rm sys}+166$ km~s$^{-1}$ 
and $V_{\rm sys}-121$ km~s$^{-1}$ ;
see figure 4 and table 3 in][]{kameno23a}. 

In addition to the features described above,  
a further blueshifted absorption feature is detected 
at a velocity range of $-500$ to $-300$ km~s$^{-1}$ 
relative to $V_{\rm sys}$. 
The full SO absorption profile spans a broad velocity range of 
$> 700$ km~s$^{-1}$ (from 1015 to 1731 km~s$^{-1}$). 
It can be decomposed into three distinct velocity groups, labeled Blue1, Blue2 and Red1 in figure~\ref{fig:socenter}. 
Blue1 and Red1 are prominent,
with peak absorption depths exceeding $ 5\%$ of the continuum level.
This depth is larger than that seen in the ALMA data  \citep[$\sim 2 \%$;][]{kameno23a}. 
The overall profile appears nearly symmetric around $V_{\rm sys}$, 
in contrast to the asymmetric profile reported in the ALMA results.

A triple-Gaussian fit was applied to the spectrum, 
resulting in a $\chi^2$ per degree of freedom  $=$ 36.8 / 29. 
The best-fitting Gaussian components are overlaid on the spectrum 
in figure~\ref{fig:socenter}, and 
their parameters are listed in table~\ref{tab:sogauss}. 
To assess the statistical significance of the fitted absorption components, 
we applied a $t$-test under the null hypothesis that 
the peak amplitude ($A_{\rm p}$) is zero. 
The resulting $p$-values were less than 0.01 for all three components, 
confirming their statistical significant. 

We examined whether  Blue1 could be attributed to SO$_2$ $10_{2,8}-10_{1,9}$ ($\nu_{\rm rest} =$ 129.51481 GHz) 
or SiO $J=3-2$, $v=1$ ($\nu_{\rm rest} =$ 129.36335 GHz). 
\citet{kameno23a} identified SO$_2$ $10_{2,8}-10_{1,9}$ absorption at $V_{\rm sys}+161$ km s$^{-1}$ in the ALMA data. 
This absorption would appear at $V_{\rm sys}-712$ km s$^{-1}$, 
which differs by 300 km~s$^{-1}$ from the velocity of Blue1 in figure \ref{fig:socenter}. 
On the other hand, the SiO $J=3-2$, $v=1$ line would be located at $V_{\rm sys}-465$ km~s$^{-1}$, which is within the velocity range of Blue1. 
However, no other SiO transitions have been found so far. 
Thus, we conclude that Blue 1 can be ascribed to blueshifted SO absorption.

\begin{table*}
  \tbl{Parameters of the SO $J_N = 3_3-2_2$ Gaussian line  components}{%
  \begin{tabular}{lcccccc}
    \hline \hline
    Absorption
    & $A_{\rm p}$\footnotemark[$*$]
    & FWHM\footnotemark[$\dag$] 
    & $V_\mathrm{ctr}$\footnotemark[$\|$] 
    & $V_\mathrm{ctr}-V_\mathrm{sys}$  
    & \multicolumn{2}{c}{$t$-test for $A_{\rm p}$}  \\
 \cline{6-7}   
    &   &   &   &   & $t$-value  & $p$-value \\ 
    &  & (km~s$^{-1}$) & (km~s$^{-1}$) & (km~s$^{-1}$) &  & \\
      \hline
    Blue1 & $-0.057\pm0.014$ & $183\pm22$  & $1095\pm22$ & $-412$ & $-4.21$ & 0.000227 \\
    Blue2 & 
  $-0.050\pm0.018$ & $80\pm32$  & $1392 \pm 21$ & $-115$ & $-2.78$ & 0.009444 \\
    Red1 & $-0.061\pm0.014$ & $173\pm46$ & $1730\pm20$ & $+223$  & $-4.32$ & 0.000166 \\
    \hline
  \end{tabular}}
\label{tab:sogauss}
\begin{tabnote} 
\footnotemark[$*$]  Peak amplitude of Gaussian component in normalized flux density.  \\
\footnotemark[$\dag$]  Full width at half depth in velocity.  \\ 
\footnotemark[$\|$]  Centroid velocity. \\
\end{tabnote}
\end{table*}

\begin{figure}
 \begin{center}
  \includegraphics[width=8cm]{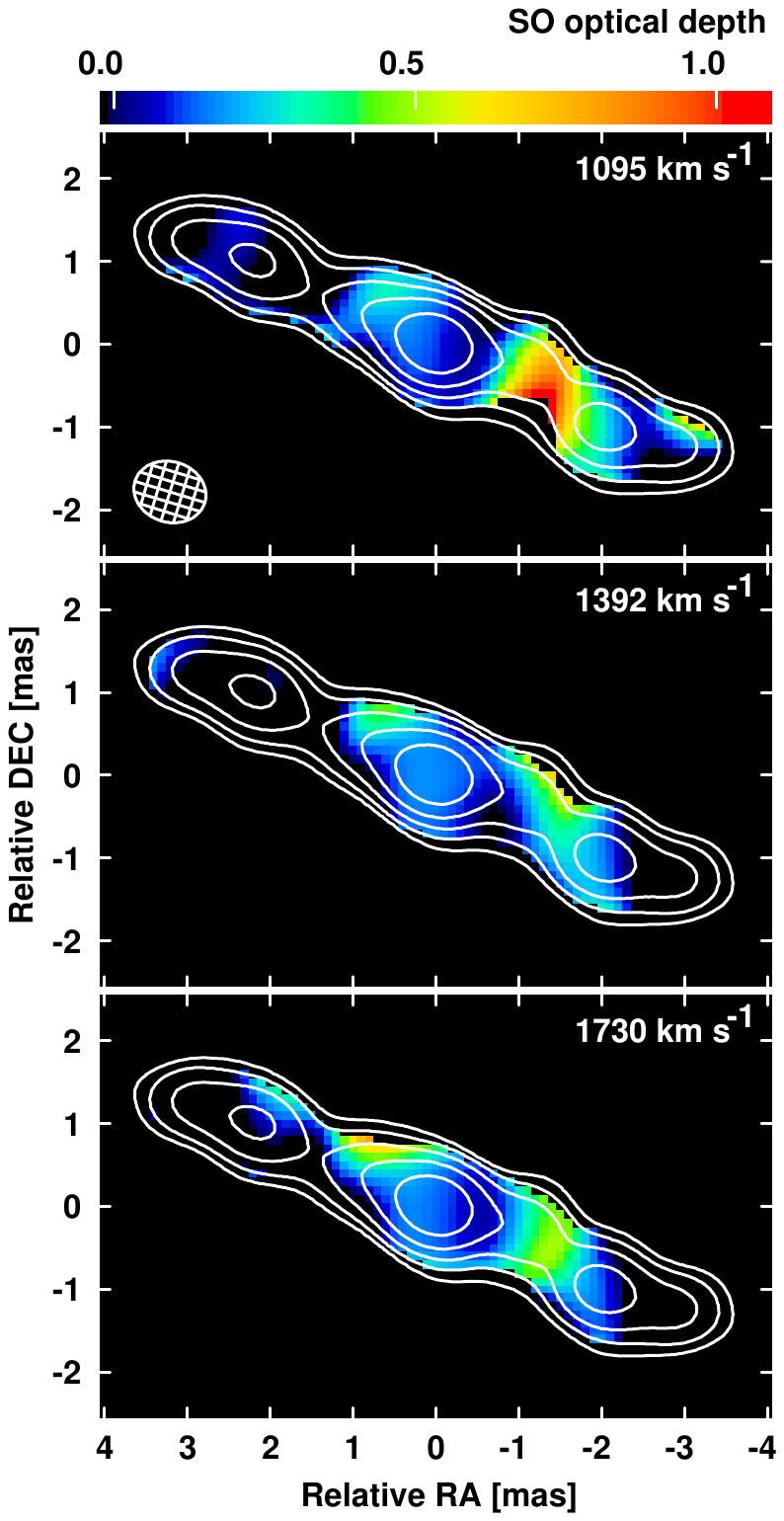} 
 \end{center}
\caption{
Color optical depth channel maps of 
three SO absorption components.  
(Top) Blue1 at 1095 km~s$^{-1}$, 
(Middle) Blue2 at 1392 km~s$^{-1}$, 
(Bottom) Red1 at 1730 km~s$^{-1}$. 
They are overlaid by the continuum contour map of NGC 1052 
at $\lambda$2 mm also shown in figure~\ref{fig:dcntmap}. 
The centroid velocity of each component is shown 
at the upper right. 
The rms noise in optical depth is typically 0.2.
{Alt text: 
Three color panels comparing the optical depth of different 
SO absorption components. 
}
}
\label{fig:soopd}
\end{figure}

Figure~\ref{fig:soopd} illustrates 
the spatial distributions of SO optical depth in absorption,  
overlaid on the continuum contour map 
(figure~\ref{fig:dcntmap}). 
The maps reveal that SO absorption predominantly occurs 
toward the central continuum components, E2, Core, and W2. 
The SO optical depth distribution 
along the central components is asymmetric,  
with a pronounced enhancement toward the receding central component W2.  
In contrast, 
the outer jet components E1 and W1 exhibit little or 
no significant absorption 
at levels above three times the rms noise. 
The optical depth of SO measured with a high spatial resolution of $< 0.1$ pc achieved by the KVN is an order of magnitude higher than that from the ALMA observations by \citet{kameno23a}.

\begin{table}
  \tbl{SO $J_N = 3_3-2_2$ optical depth for each absorption component}{%
  \begin{tabular}{lcccc}
    \hline \hline
    Absorption
    & $\tau_{\rm max}$
    & $\langle \tau \rangle$
    & $f_c$ 
    & $\tau_{\rm 30pc}$ \\
(1) & (2) & (3) & (4) & (5) \\
\hline
    Blue1 & 1.078 & 0.133 & 0.123 & -- \\
    Blue2 & 0.593 & 0.065 & 0.110 & 0.010 \\
    Red1  & 0.508 & 0.075 & 0.147 & 0.021 \\
    \hline
  \end{tabular}}
\label{tab:tau}
\begin{tabnote} 
Col.(1) Absorption ID. 
Col.(2) Maximum value of optical depth against W2.
Col.(3) Mean optical depth. 
Col.(4) Mean covering factor. $f_c = \langle \tau \rangle / \tau_{\rm max}$.  
Col.(5) Optical depth  measured with a 30 pc spatial resolution by ALMA, from figure 3 in \citet{kameno23a}. 
\end{tabnote}
\end{table}

\section{Discussion}

\subsection{Continuum emission properties of the central source}

The total 129 GHz continuum flux density of the central source (E2 $+$ Core $+$ W2) was measured to be 220 mJy in 2019. 
Assuming that the flux densities remained stable from 2017 to 2019, and using the 129 GHz flux density in 2019 together with the 89 GHz continuum flux density of the central component within 1.6 mas obtained in 2017 \citep[414 mJy;][]{sss19}, 
we derive a steep spectral index 
$\alpha = -1.7$ 
($S_{\nu} \propto \nu^{+\alpha}$) 
over the 89--129 GHz frequency range
for the central source. 
As only two frequency data points are available, 
 it remains uncertain 
whether the turnover frequency lies above or below 89 GHz.
Nevertheless, 
we can conclude 
that the central source has 
an optically thin synchrotron spectrum at 129 GHz. 

The lower limit of the brightness temperature 
($T_{b}$) of the central source is estimated 
as > $10^{7}$ K at 129 GHz, 
using the following formula 
\begin{equation}
T_{b} = 2 \ln 2 
\frac{c^2 S_{\nu}}{\pi k_{\rm B}r^2 \nu^2} 
(1+z)
\end{equation}
where $c$ is the light speed, 
$S_{\nu}$ is the flux density at the frequency $\nu$, 
$k_{\rm B}$ is the Boltzmann constant, 
$r$ is the radius of the component and 
$z$ is the redshift. 
The obtained $T_{b}$ value supports the above argument 
that the detected continuum emission 
is dominated by non-thermal synchrotron emission.

\subsection{SO absorption line profile and optical depth}

As described in subsection~\ref{sec:so}, 
the SO $J_N = 3_3-2_2$ absorption 
in the KVN results 
is deeper than that of the ALMA data.  
The observed difference in depths may be mainly due to 
the different spatial resolution 
between the KVN ($\sim0.7$ mas, corresponding to 0.07~pc) and  ALMA ($\sim0.\!\!^{\prime\prime}3$, corresponding to 30~pc) observations. 
On sub-arcsecond scales with  ALMA, 
the background continuum emission with the extended jets covers a region of several tens of parsecs, whereas the SO gas surrounding the SMBH is distributed within a much more compact region. 
Consequently, the covering factor of the gas becomes less than unity as the resolution becomes coarser.

Here we examine the effect of resolution on optical depth by estimating the intensity-weighted mean optical depth 
$\langle \tau \rangle$ 
and the mean covering factor 
 $f_c = \langle \tau \rangle / \tau_{\rm max}$ 
for the absorption components over the whole two-sided nuclear jet structure of the sub pc scale region (figure~\ref{fig:soopd}) in the same manner as that adopted by \citet{sss16} and \citet{kameno20}. 
We estimate 
the mean optical depth $\langle \tau \rangle$ 
for each SO absorption feature,   
using
\begin{equation}
    \langle \tau \rangle 
    = 
    \frac{\iint \tau(x,y) I (x,y) dx dy}
    {\iint I(x,y) dx dy}, 
\end{equation}
where $\tau (x,y)$ is the optical depth distribution 
of absorption feature 
shown in figure~\ref{fig:soopd},
and $I(x,y)$ is the 129-GHz continuum image 
shown in figure~\ref{fig:dcntmap}. 
The estimated $\langle \tau \rangle$ and $f_c$ are listed 
in table~\ref{tab:tau},  
together with the SO $J_N = 3_3-2_2$ 
optical depth measured with a 30~pc spatial resolution by  ALMA ($\tau_{\rm 30pc}$) from figure 3 in \citet{kameno23a}. 
The derived mean covering factor $f_c$ values are 
slightly lower than the covering factor yielded from the HCN $J=3-2$ absorption \citep[0.17;][]{kameno23a}. 
However, the values of $\langle \tau \rangle$ are at least three times higher than $\tau_{\rm 30pc}$. 
The discrepancy in optical depth can be 
attributed to differences in spatial resolution. Specifically, continuum emission from the extended jet of NGC 1052 is detected with ALMA but resolved out in the KVN data. 
Moreover, contamination from SO $J_N = 3_3-2_2$ emission may partially fill the absorption in the ALMA data, 
while the KVN spatially excludes such thermal contamination. 
It should  also be noted that \citet{kameno23a} calculate 
the contamination-corrected 
SO $J_N = 3_3-2_2$ optical depth 
under the assumption that the extra line equivalent width is equal to the average of other SO$_2$ lines at 128.6 and 129.5 GHz, while the SO $J_N = 3_3-2_2$ optical depth derived from our KVN observations might be the result of possible blending with the SO$_2$ $12_{1,11}-11_{2,10}$ line.

\subsection{Location and kinematics of SO absorption}

The absence of significant SO absorption against the outer jet components 
E1 and W1 places a spatial constraint on the distribution of the SO gas.  
The spatial extent of the SO gas must therefore be smaller than the angular separation between E1 and W1 (4.8 mas), which corresponds to an apparent size of less than 0.45 pc on the sky.
The detection of absorption only toward the central source, and not toward the outer receding component W1, contrasts with the characteristics observed in the 
$\lambda$3mm HCN and HCO$^{+}$ absorptions  \citep{sss16,sss19}. 
This concentration toward the central source is inconsistent with the SO gas being located in the molecular layer at the outer radii of the torus,   
where the HCN and HCO$^{+}$ gases lie. 
The enhanced SO optical depth 
of the central receding component W2 
suggests that 
the SO gas resides 
at or close to the inner edge ($\ll$ 1 pc)
of the near side of the torus. 
The compact distribution of the high SO optical depth 
seen toward W2 at multiple velocities 
implies that 
small ($<$ 0.1~pc) individual SO gas clumps 
are moving at various velocities 
across the line of sight to W2, 
which lies in the immediate vicinity of the SMBH. 
The variations in SO optical depth across the velocity channel maps (figure~\ref{fig:soopd}) highlight the inhomogeneous nature of the absorbing gas, supporting a clumpy gas structure. 
Since the velocity of the Red1 component is close to 
the peak velocities of 22 GHz H$_2$O maser 
emission and HCN $J=$ 1--0 absorption, 
it is likely that 
Red1 arises from the ongoing infalling gas 
inside the torus toward the SMBH. 
This interpretation is in agreement with earlier  results 
for the redshifted H$_2$O maser  and 
HCN absorption features \citep{sss08, sss16}. 
On the other hand, 
Blue1 and Blue2 are the signature of at least two 
individual outward motions in the region.

The derived 129 GHz SO absorption spectrum and maps show partial similarities to the 321 GHz H$_2$O maser 
\citep{kameno24}
in terms of radial velocity and spatial distribution.  
In particular, several commonalities are observed between the Blue2 component of 129 GHz SO and the B1 component in the 321 GHz H$_2$O maser.
First, the velocity of Blue2 in 129 GHz SO agrees well with that of the B1 component in the 321 GHz H$_2$O maser,  
though 
the entire velocity range of SO is much wider 
than that of H$_2$O. 
Second, 
both 129 GHz SO absorption and 321 GHz H$_2$O maser distributions extend westward along the jet direction for approximately 2 mas from the continuum peak position, where the SMBH is presumed to be located. 
The 321 GHz H$_2$O maser spots of B1 are 
predominantly distributed within 1 mas 
from the optically-thin continuum peak position. 
In the case of 129 GHz SO, 
Blue1, Blue2, and Red1 are all concentrated on W2. 
Among them, 
Blue2 shows a somewhat prominent distribution 
that extends to Core as well, 
in contrast to Blue1 and Red1. 
This suggests that Blue2 of 129 GHz SO and B1 with its 321 GHz H$_2$O maser likely trace the same outward motion. 
Notably, however, 
no counterpart to the 129 GHz SO-Blue1 source is found in the 321 GHz H$_2$O maser line. 
The 129 GHz SO absorption is detected toward the millimeter  jet components, 
while the 321 GHz H$_2$O maser is confined to a location in front of a compact sub-millimeter continuum source. 
Consequently, SO is more sensitive to the kinematic influence of the interaction region compared to H$_2$O, 
which could explain why the outflow traced by Blue1 is detectable only in SO.
The spatial coincidence between 129 GHz SO absorption and the 321 GHz H$_2$O maser implies that the H$_2$O maser could be excited by the jet-torus interaction.

\begin{figure}
 \begin{center}
  \includegraphics[width=8cm]{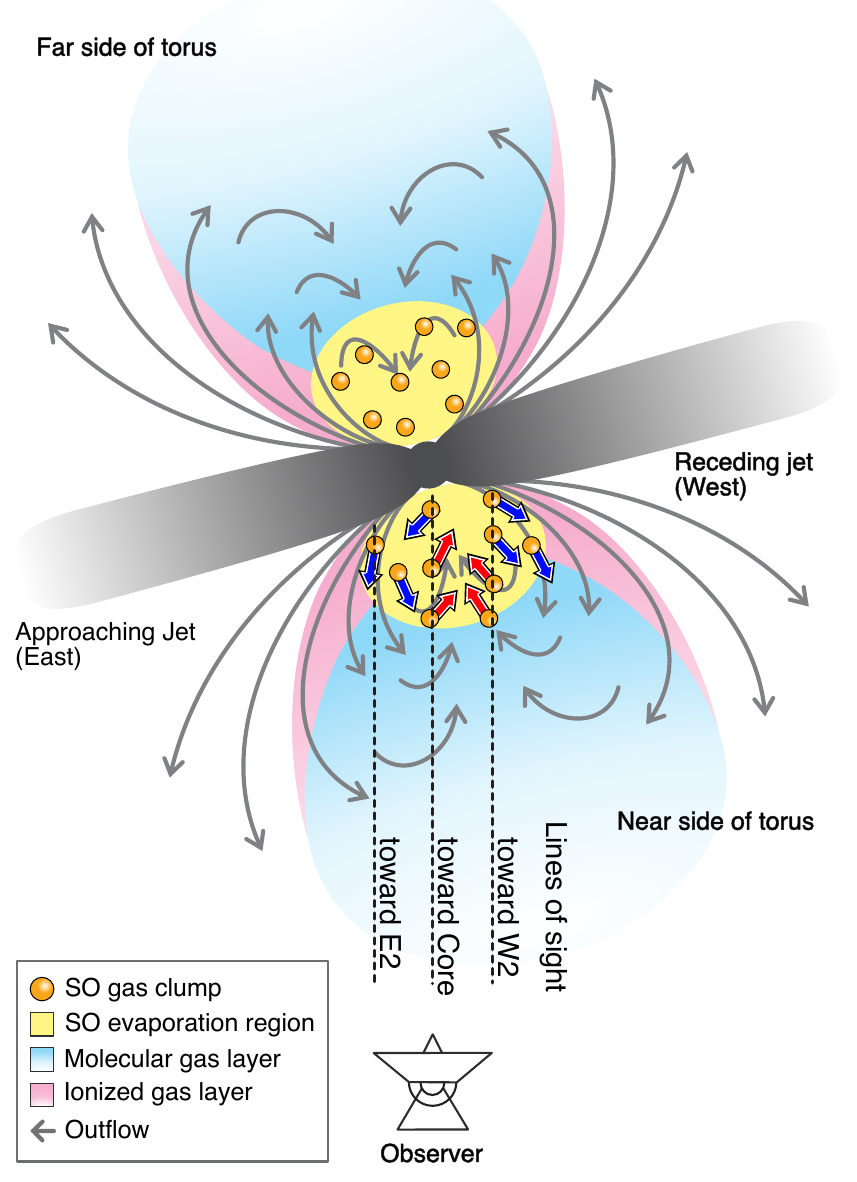} 
 \end{center}
\caption{
A schematic view of the sub-pc nuclear region in NGC 1052. 
The two-sided jet axis is inclined by $\ge 76^{\circ}$  with respect to the line of sight \citep{sss08}.
The torus has several phase layers of plasma and molecules. 
Shock heating caused by interactions between sub-relativistic jets and torus gas generates a compact SO evaporation region, smaller than 0.45 pc.
Clumps containing evaporated SO are carried outward by the jet, forming outflows (blue arrows). 
Some of the clumps fall back onto the 
equatorial plane and eventually infall toward the SMBH (red arrows). 
The longer line-of-sight path through the SO-rich region toward the central receding component explains its high optical depth. 
{Alt text: 
A cartoon showing the possible environment in the 
torus and jets in NGC 1052. 
}
}
\label{fig:somodel}
\end{figure}

We have developed a plausible multiphase torus model 
(figure~\ref{fig:somodel}) 
that accounts for the observed characteristics of 
the SO absorption. 
At the innermost radii of the torus, 
the SO evaporation region is formed through  shock heating
resulting from the interaction between 
the sub-relativistic jets and torus gas, 
as suggested by \citet{kameno23a}. 
The apparent size of the SO evaporation region is more compact than 0.45 pc. 
The evaporated SO molecular clumps are transported 
downstream by the jet, inducing an outflow. 
A fraction of the outflowing clumps subsequently falls  back 
onto the equatorial plane of the torus 
and 
eventually infalls toward the SMBH through the torus. 
The infall velocity is considered to be approximately 200 km~s$^{-1}$. 
The lines of sight to the central components intersect 
both outflowing and infalling clumps.  
Since the jet axis is slightly inclined relative to the sky plane, 
the near side of the torus should mainly obscure the SMBH 
and the receding jet side.
The central receding component W2 exhibits a longer path length through the SO evaporation region along the line of sight compared to the approaching component E2.
The difference in path length explains the high 
optical depth concentration observed toward W2.
The model indicates that both outflow and infall coexist in a clumpy and dynamic structure within the sub-parsec region of NGC 1052, consistent with a clumpy dynamic torus 
 \citep[e.g.,][]{wada09,kudoh23}.

NGC 1052 represents the first case in which sub-pc-scale jet–torus interactions, along with associated dynamical phenomena,  
such as inflows, outflows, and shock-induced molecular gas excitation,
have been explored via sulfur-bearing molecular absorptions.
Further applications of this observational approach to a broader AGN sample will contribute to a better understanding of the detailed physical processes governing jet–torus interactions and their impact on the circumnuclear environment.

\section{Conclusions}

We present the first sub-pc scale images of continuum 
and SO absorption in the center of NGC 1052 
at 129 GHz  using the KVN. 
The continuum image 
reveals a bright central source and symmetric two-sided jets extending 0.75 pc, 
with five distinct components (E1, E2, Core, W2, W1)
identified. 
These components are aligned along a position angle of 64$^{\circ}$, 
consistent with prior lower-frequency VLBI studies. 
The central source exhibits a steep spectral index 
($\alpha$ = –1.7) between 89 and 129 GHz, indicating optically thin synchrotron emission at 129 GHz. 
Its brightness temperature exceeds $10^7$ K, further supporting a non-thermal origin.
SO absorption is clearly detected toward the central components Core, W2 and marginally at E2, with no significant absorption at E1 and W1. 
The absorption profile spans a broad velocity range over 700 km~s$^{-1}$, showing three distinct velocity groups. 
The SO absorption is concentrated toward the central components, with a pronounced enhancement at the receding W2 component. 
The spatial distribution of SO absorption indicates that the absorbing gas is concentrated within a region smaller than  0.45 pc, likely located close to the inner edge of the torus. 
The distribution and velocity structure of SO absorption suggest that the SO gas forms clumpy and small-scale ($<$ 0.1 pc) structures, showing both inflow and outflow motions. 
To explain the spatial and kinematic properties of the SO absorption, 
we propose 
a multiphase torus model in which  
jet-torus interactions produce shock-heated regions at the inner edge of the torus, leading to SO gas evaporation and the formation of SO clumps.
These clumps are entrained by the jet and transported outward, while some clumps fall back toward the SMBH, creating inflow signatures. The line of sight intersects both outflowing and infalling clumps, producing the observed absorption profile and velocity structure. 
Our observations also reveal that 
the 129 GHz SO absorption shows partial similarities to the 321 GHz H$_2$O maser in both radial velocity and spatial distribution.
The spatial coincidence between SO absorption and H$_2$O maser emission suggests that the excitation of the H$_2$O maser could be linked to the jet–torus interaction.  


\begin{ack}

We are grateful for the excellent support provided 
by the staff of the KVN.  
The KVN and a high-performance computing cluster are facilities operated by the KASI (Korea Astronomy and Space Science Institute). The KVN observations and correlations are supported through the high-speed network connections among the KVN sites provided by the KREONET (Korea Research Environment Open NETwork), which is managed and operated by the KISTI (Korea Institute of Science and Technology Information).
This research was partially supported by JSPS KAKENHI grant No. 21K03628.
\end{ack}

\end{document}